\documentclass[prc,aps,epsfig,twocolumn,nofootinbib]{revtex4}

\usepackage{graphicx} 

\begin{document}

\date{\today}
\title{
Probing pre-formed alpha particles in the ground state of nuclei.}

\author{J.A. Scarpaci$^1$, M. Fallot$^{1,*}$, D. Lacroix$^2$, M. Assi\'e$^1$, L. Lefebvre$^1$, N. Frascaria$^1$, D. Beaumel$^1$, C. Bhar$^1$,  
Y. Blumenfeld$^1$, A. Chbihi$^2$, Ph. Chomaz$^{2,\dagger}$, P. Desesquelles$^{1,**}$, J. Frankland$^2$, H. Idbarkach$^1$,
E. Khan$^1$, J.L. Laville$^2$, E. Plagnol$^{1,***}$, E.C. Pollacco$^3$, P. Roussel-Chomaz$^2$, J.C. Roynette$^1$, 
A. Shrivastava$^{1,****}$, T. Zerguerras$^1$}
\affiliation{$^1$Institut de Physique Nucl\'eaire, CNRS/IN2P3, Universit\'e Paris-Sud 11, F-91406 Orsay, France}
\affiliation{$^2$GANIL, CEA/DSM - CNRS/IN2P3, Bd Henri Becquerel, BP 55027, F-14076 Caen Cedex 5, France}
\affiliation{$^3$CEA-Saclay, DSM/IRFU SPhN, F-91191 Gif sur Yvette Cedex, France}

\begin{abstract}
In this Letter, we report on alpha particle emission through the nuclear break-up in the reaction 
$^{40}$Ca on a $^{40}$Ca target at 50A MeV. It is observed that, similarly
to nucleons, alpha particles can be emitted to the continuum with very specific angular distribution 
during the reaction. The alpha particle properties can be understood as resulting from an alpha 
cluster in the daughter nucleus that is perturbed by the short range nuclear attraction 
of the collision partner and emitted. 
A time-dependent theory that describe the alpha particle wave-function evolution is able to
reproduce qualitatively the observed angular distribution. This mechanism offers new 
possibilities to study alpha particle properties in the nuclear medium.   
\end{abstract}
\maketitle


Nuclei are complex self-bound systems formed of nucleons. Conjointly to a mean-field picture where 
nucleons can be regarded as independent particles, few nucleons 
might self-organize into compact objects, called clusters, inside the nucleus. 
The understanding of clustering in the nuclear medium is a central issue and has been the subject to 
extensive research. Clustering in $N=Z$ nuclei has a long standing 
history (see discussion in \cite{Von06}) and recent AMD calculation are studying the $\alpha$+$^{36}$Ar configuration of the
$^{40}$Ca nucleus \cite{tan07} for which experimental results are presented here. 
The states based on $\alpha$ particles have also been observed in (d,$^6$Li) 
experiment on $^{40}$Ca
\cite{ume84}, however they are usually not so much found in the ground 
states (GS) but rather 
observed as excited states close to the decay thresholds into clusters, as suggested by Ikeda \cite{ike72}. In particular, 
the Hoyle state \cite{che07}, i.e., the 0$^+_2$ state at 7.65 MeV in $^{12}$C which was interpreted as an alpha-particle
condensate \cite{toh01}, and other similar states in heavier n$\alpha$ nuclei, have attracted much renewed 
attention (see, e.g., \cite{sch07}). Other experiments have observed the emission of alpha particles but with 
different mechanisms 
such as the decay of giant resonances \cite{fal05-2} or the emission from the neck \cite{hud04}.

In this Letter we report on sudden $\alpha$ particle emission resulting of the 
nuclear break-up, also called the "Towing mode", of $\alpha$ cluster present in the 
{\it ground state} of $^{40}$Ca. This mechanism, first observed 
with neutrons and protons in the inelastic scattering channels with stable beams \cite{Sca98} is an emission to the continuum 
due to the nuclear potential of the passing by nucleus. The time scale is much shorter here than for a regular 
evaporation as it is related to passing time of the projectile of the order of 10$^{-22}$ s.
It was understood with a dedicated Time Dependent Schr\"odinger Equation (TDSE) \cite{Lac99} technique as follows. 
As the emitter nucleus passes close to the collision partner, the short-range nuclear potential of the latter 
attracts the least bound nucleons leading to the emission of particles at mid-rapidity. We have shown an excellent 
reproduction of the experimental properties with the model calculation and we have demonstrated that this mechanism 
has several interesting aspects. The characteristics of the emission (angular and energy distributions) strongly 
depend on the initial quantum 
properties of the towed particle (angular momentum, extension of the wave-function and binding energy). 
These properties promote the "Towing mode" as a tool of choice to infer spectroscopic information of nucleons in nuclei as 
shown
in ref. \cite{Lim07,Ass08,Ass09} for $^{11}$Be and $^6$He. 

In this present experiment, it has been observed not only 
the target break-up of one proton \cite{fal05}, but also of $\alpha$ particle \cite{fal05-2} as we will present 
in the following, indicating the formation 
of $\alpha$ cluster in the $^{40}$Ca nucleus. The $\alpha$ cluster is emitted to the continuum due to the 
perturbation of the other nucleus with specific angular and energetic properties. A dedicated model describing the evolution 
of the $\alpha$ particle wave-function initially in the daughter nucleus for grazing impact parameters is able to 
describe qualitatively the observed behavior.

The experiment was performed at the GANIL facility, by bombarding a self-supported 0.2 mg/cm$^2$ natural Ca target with a 
50A MeV $^{40}$Ca beam. The ejectiles were identified in the focal plane of the SPEG spectrometer \cite{bia89}, 
in coincidence with the light 
charged particles detected in 240 CsI(Tl) detectors of the INDRA $4\pi$ array \cite{pou95} covering an angular range from 
14$^\circ$ to 176$^\circ$ with respect to the beam direction. The  projectiles were measured between 
$\theta_{lab}=0.5^\circ$ and $5^\circ$ using the SPEG spectrometer with its standard detection system. Stripped cathode 
drift chambers were used for the position measurement in the focal plane \cite{dro00}. Combined with 
the very thin target, they allowed for an excitation energy resolution of 350 keV, 
i.e., $\Delta p/p $= 9 $\times$ 10$^{-5}$. The ejectiles were unambiguously identified in charge and mass through energy 
and time of flight measurements and in this letter we will concentrate on the inelastic channel only where 
the ejectiles are of the same nature as the $^{40}$Ca projectile.

\begin{widetext}

\begin{figure}
\begin{center}
\includegraphics[width=14.cm]{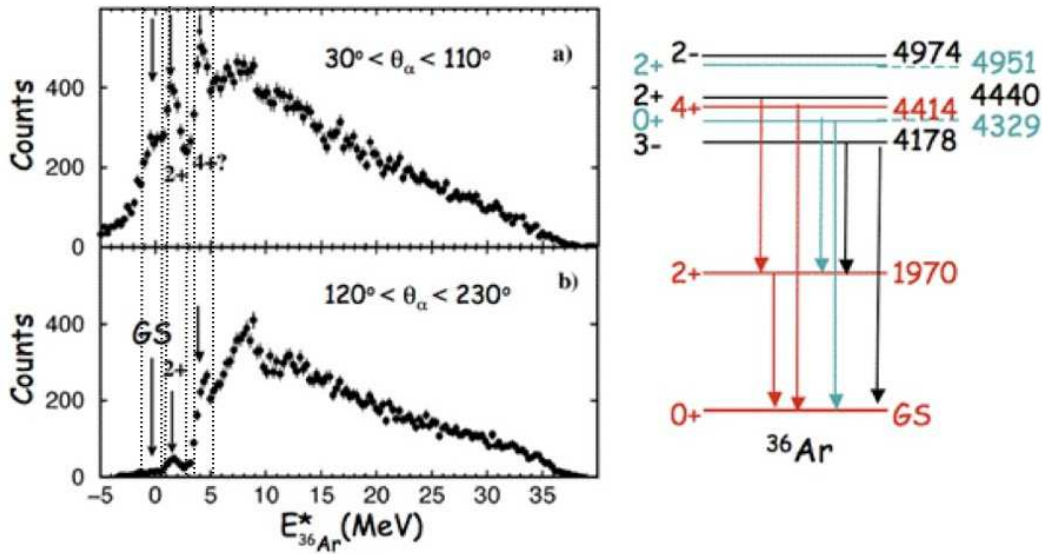}
\caption{(Color online) Missing energy spectrum of $^{36}$Ar in the reaction $^{40}$Ca($^{40}$Ca,$^{40}$Ca+$\alpha$) for an excitation
energy range between 20 and 50 MeV. Fig. (a): $\alpha$
particles are detected between 30 and 110 degrees. (b): $\alpha$ particles are detected between 
130 and 220 degrees in the laboratory system, 
in the backward direction compared to the beam. The arrows show the GS, the 2$^+$ state and a state around 4.4 MeV. 
The dashed lines are the upper and lower limits of the gates used to extract the angular distributions presented in 
Fig.\ref{fig2:alpha}. Lower and higher values for the gates were as follows: from -1.1 to 0.7 MeV to select most of 
the GS ; from 1.1 to 2.9 MeV 
for the 2$^+$ state and from 3.7 to 5.1 for state around 4 MeV. The Right panel shows the level scheme of known states 
up to 4.974 MeV excitation energy in $^{36}$Ar \cite{nndc}.} 
\label{fig1:alpha}
\end{center}
\end{figure}

\end{widetext}

The identification of the particles was obtained by analyzing the fast and slow components of the CsI 
scintillator light, and was fully reliable above particle kinetic energy of 4 MeV. We have therefore set a 4 MeV 
software energy threshold for the $\alpha$ particles. The light 
particle energy calibration was performed using 
the particle decay towards the GS 
of the daughter nucleus \cite{sca97,fal05}. 
The obtained energy resolution for $\alpha$ particles 
is 1.1 MeV hence a 1.15 MeV energy resolution on the reconstructed missing energy (see below).

We performed a measurement of the missing energy calculated as follows: 
$E_{\rm miss} = E^*_{\rm Ca} - E^{\rm CM}_{\alpha} - E_{\rm recoil}(^{36}{\rm Ar})$
where $E^*_{\rm Ca}$ is the initial excitation energy in $^{40}$Ca obtained by the measurement of the inelastically scattered projectile 
detected after the SPEG spectrometer, $E^{\rm CM}_{\alpha}$ the $\alpha$ particle energy 
in the center of mass frame of the recoiling $^{40}$Ca target, and $E_{\rm recoil}$ is the recoil energy of target-like $^{36}$Ar. 
The remaining excitation energy of $^{36}$Ar is then E$^{*}_{^{36}Ar}$ = E$_{\rm miss}$ - Q$_\alpha$ and is plotted
in Fig.\ref{fig1:alpha}. We show the final states of $^{36}$Ar after an $\alpha$ 
particle was removed from the target and detected in two different angular regions (for an apparent excitation
energy range between 20 and 50 MeV). Spectra (a) and (b) correspond to forward and backward emission respectively. 
Region (b) is selecting the evaporation 
of an excited $^{40}$Ca target that shall emit isotropically. This statistical decay is then also present in 
Fig.\ref{fig1:alpha} (a) with on top a strong 
feeding of several low energy states of $^{36}$Ar. On spectrum (a), we clearly recognize the 0$^+$ GS, 
the first 2$^+$ state at 1.97 MeV and a second excited state around 4.4 MeV which
could not be separated from the surrounding states (see right panel of Fig.\ref{fig1:alpha}) 
but could possibly be assigned to the 4$^+$ state of the $^{36}$Ar.  
On spectrum (b), these states are much less populated. 
The difference observed between the forward and backward $\alpha$ particle emission is very similar to 
the anisotropy in nucleon emission measured previously in reaction $^{58}$Ni($^{40}$Ar,$^{40}$Ar+p or n)$^{57}$X 
(see Ref.\cite{Sca98}) which is a clear signature of nuclear break-up, called 
sometime "Towing mode" where the target after its break-up is left in its GS or with very small excitation energy. Because of 
this sudden nuclear break-up, the $\alpha$ particle is expelled from a cold nucleus, thus the idea of a pre-formation in 
the GS of this nucleus.

This first evidence of a similar phenomenon with $\alpha$ particles motivates the precise understanding of this observation in 
order to provide a new tool for the study of $\alpha$ particle properties in the nuclear medium. Extrapolating results obtained 
with nucleons, we do expect that the properties of the emitted particles depend significantly on the initial $\alpha$ particle 
properties in the emitter (binding energy, quantum numbers, wave-function extension). 
Here, we will follow exactly the same strategy as for single nucleon emission and question if 
the observed anisotropy in $\alpha$ angular distribution can be understood as the nuclear break-up of a preformed $\alpha$ 
particle wave-function.

The mechanism responsible for anisotropic distribution of $\alpha$ particles seems to feed different states of a band corresponding 
to a deformed $^{36}$Ar nucleus with a $\beta$ of 0.256 \cite{ram01} (see Fig.\ref{fig1:alpha} right).
The feeding of several states of the $^{36}$Ar can eventually be understood as follows. Assuming that the 
0$^+$ GS of the $^{40}$Ca presents 
different configurations corresponding to an $\alpha$ particle wave-function coupled 
to an $^{36}$Ar core, eventually excited, we write a configuration as $(| \varphi_\alpha \rangle \otimes | (^{36}$Ar$)^* \rangle)_n$ 
where $n$ labels the configuration while $\varphi_\alpha$ and $| (^{36}$Ar$)^* \rangle$ 
denote respectively a specific $\alpha$ particle wave-function and a specific state of the $^{36}$Ar.
As the $\alpha$ particle is emitted, the $^{36}$Ar is left in the specific configuration 
$| (^{36}$Ar$)^* \rangle$ and can eventually decay through $\gamma$ emission.
Assuming that the $\alpha$ configurations present in the $^{40}$Ca GS do not interfere with each others, 
the feeding of different states can directly be assigned to the independent contributions of each configuration.
A complete understanding of the feeding process and/or the extraction of $\alpha$ particle spectroscopic factors would require 
very good $\alpha$ particle energy resolution as well as $\gamma$ coincidence. 
Due to absence of gamma detection and insufficient energy resolution (1.15 MeV) of the missing energy spectrum 
(Fig.\ref{fig1:alpha}), separation between the GS and the first 2$^+$ state, and the states around 4 MeV in 
$^{36}$Ar could not be precisely achieved. However, in spite of these limitations, the angular cross-section 
of the emitted $\alpha$ 
particles could be extracted. 
Here, the compatibility of the scenario described above will be tested qualitatively on the angular distributions of the $\alpha$ 
particles associated with a particular state in the $^{36}$Ar against the experimental observation.  

\begin{figure}
\includegraphics[width=8.cm]{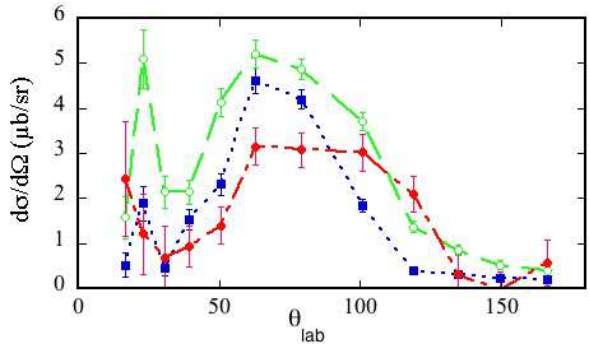}
\caption{(Color online)
Angular cross-sections of $\alpha$ particles when the three first states around $E^*_{^{36}{\rm Ar}} \simeq 0$ (blue dotted line), 
$2$ (green dashed line) and $4$ MeV (red dot-dashed line) 
in figure \ref{fig1:alpha} are fed respectively. For the later state, a subtraction was made with the distribution obtained for  the 8 to 9 MeV region.
}
\label{fig2:alpha}
\end{figure}

In order to isolate and characterize the contribution of a given configuration, the experimental $\alpha$ particle 
angular distributions were extracted by setting gates around the GS, the first 2$^+$ state and the state about 4 MeV,
bearing in mind that this selection allows for some contribution from one state to the others. 
These gates are presented on the left spectra of Fig.\ref{fig1:alpha} (vertical dashed lines) 
and the corresponding distributions are displayed in Fig.\ref{fig2:alpha}. For the state around 4 MeV, 
as it has some contribution from evaporation as seeing in Fig.\ref{fig1:alpha} b), a subtraction was made 
of the 8 to 9 MeV region which only decays through evaporation and that shows a rather flat contribution (not shown here). 
A clear sensitivity of the angular distribution to the final state is observed in this figure. 

To test if these distributions could be understood as the final product of an $\alpha$ particle initially in the $^{40}$Ca 
and emitted to the continuum as the 
projectile passes by, our TDSE technique was applied to the emission of alphas.
In this model, an $\alpha$ particle wave-function, denoted by $\varphi_\alpha$ is first 
initialized in a spherical symmetric potential describing the core-alpha
interaction. The potential of ref.\cite{Buc95}, which has been optimized 
for nuclei in the same mass region to reproduce nuclear structure spectra, $B(E_2)$ transition strength, $\alpha$ decay widths 
and elastic scattering cross-sections, is used here. Its nuclear part is given by 
\begin{eqnarray}
V(r) = V_0 \left\{ \frac{\alpha}{\left(1+e^{(r-R_0)/a_0}\right)} + \frac{1-\alpha}{\left(1+e^{(r-R_0)/a_0}\right)^3}\right\}
\label{eq:pot}
\end{eqnarray}
where the parameters are taken as $V_0=-175.7$ MeV, $a_0=0.73$ fm, $\alpha=$ 0.3, and $R_0=$ 4.33 fm and is complemented
by a Coulomb repulsive potential. 
As stressed in ref.\cite{Buc95}, to approximately account for the Pauli blocking effect coming from the presence of
nucleons in the core, only levels with $2n+L$ $\ge$ 12, with $n$ and $L$ being respectively the radial and angular 
quantum numbers, are considered. For $^{40}$Ca, the first states 
that the $\alpha$ particle might occupy are the 6s, 5d and 4g 
and have respective binding energies of 7.0, 6.4 and 5.5 MeV.  

Starting from one of this $\alpha$ particle wave-function, the corresponding nuclear break-up has been 
studied by solving the Schr\"odinger equation:
\begin{widetext}
\begin{eqnarray}
i\hbar \frac{d}{dt} \varphi_\alpha ({\mathbf r},t)  &=& \left\{ \frac{{\mathbf p}^2}{2m_\alpha} + 
V_{\alpha {\rm Ar}}({\mathbf r}-{\mathbf R}_T(t)) + 
V_{\alpha {\rm Ca}}({\mathbf r}-{\mathbf R}_P (t))\right\}  \varphi_\alpha ({\mathbf r},t) 
\end{eqnarray}
\end{widetext}
where $m_\alpha$ is the mass of the $\alpha$ particle, 
$V_{\alpha {\rm Ar}}$ and $V_{\alpha {\rm Ca}}$ correspond to the target and projectile mean-field potentials respectively 
(both taken identical to (\ref{eq:pot})). 
The target and projectile center of mass evolution, ${\mathbf R}_T(t)$ and ${\mathbf R}_P(t)$ are 
chosen to describe a Coulomb trajectory for the passing $^{40}$Ca projectile. The TDSE is solved on a 3D mesh of 
size 259 in $\it{x}$ and $\it{y}$ ($\it{y}$ being the direction of the beam) and 199 in $\it{z}$ 
with a r-step and time step of 0.2 fm and 1.22 fm/c respectively.

\begin{figure}
\includegraphics[width=8.cm]{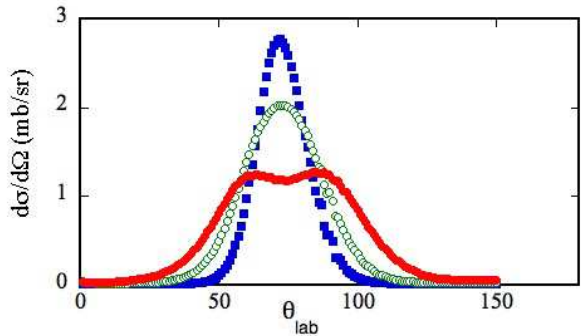}
\caption{(Color online) Calculated angular distributions for $\alpha$ particle initially in an 6s (blue squares), 
5d (green open circles) and 4g (red dots) states in $^{40}$Ca and performed with the TDSE calculation.}
\label{fig3:alpha}
\end{figure}

Similarly to the nucleon case, part of the wave-function that was initially bound by the core is emitted to 
the continuum. From the wave-function in momentum space, energetic and angular properties of emitted 
$\alpha$ can be calculated. The final angular distributions for an $\alpha$ particle wave function initially in the
6s, 5d and 4g states are shown in Fig.\ref{fig3:alpha} for an impact parameter between 9.5 and 11.5 fm.  

These calculations can be compared to the experimental angular distributions that 
leave the $^{36}$Ar nucleus in its GS and the two observed excited states presented in Fig.\ref{fig2:alpha}. 
Although core+alpha potential used in the calculation is very schematic, it is quite amazing
to see the resemblance as for the widths between the three TDSE calculations of the 6s, 5d and 4g wave functions and 
the experimental data gated respectively on the three peaks at $E^*_{^{36}{\rm Ar}} \simeq 0$, $2$ and $4$ MeV 
shown in Fig.\ref{fig2:alpha}. In particular, no adjustment was made to the nuclear potential nor 
was the possible deformation of the nucleus in which the $\alpha$ particle sits included.
In addition, as stressed above, due to $\alpha$ particle energy resolution, it is difficult to separate the different 
states and some sizable contribution of the neighboring states could be present in the angular distribution extracted 
for a specific state.

Similarly to experimental data, one could obtain absolute cross sections from the 
calculations and deduce spectroscopic factor (SF) by comparing 
theoretical and experimental results. However a precise estimate
of SF requires a perfect knowledge of the model parameters and specially of the projectile potential which is beyond the 
purpose of this letter.

Nevertheless, these findings are very encouraging: (i) they confirm the "Towing mode" scenario for the anisotropic 
emission of $\alpha$ particle recently observed (ii) it shows that experimental observation can be described 
assuming a preformed $\alpha$ particle wave-function in the GS of $^{40}$Ca. (iii) A qualitative description of the 
experiment is achieved by supposing different contributions coming from different l value of the outgoing $\alpha$ 
particle.     

Experimental evidence was shown for the emission of $\alpha$ particles during the break-up of a $^{40}$Ca target. 
A first comparison was made with a TDSE calculation assuming a preformed $\alpha$ particle in the nucleus and the 
general behavior of the measured distribution was reproduced. 
This finding opens new perspectives for the study of pre-formed $\alpha$ 
clusters in the nuclear medium. Similarly to the nucleon case, $\alpha$ particles emitted through the nuclear break-up 
channel have properties that depends sensibly on the initial wave-function (see for instance Fig. \ref{fig3:alpha} 
for the quantum number dependence). 
Following the same strategy as in the nucleon case, one might with dedicated experiments, and improved calculations, be able to access pre-formed $\alpha$ particle wave-function properties in the ground state of the nuclei.
  
$\dagger$Present address: CEA / Irfu - Centre de Saclay, F-91191 Gif sur Yvette Cedex, France

 *Present address: SUBATECH (CNRS/IN2P3 - Univ. of Nantes - EMN) F-44307 Nantes Cedex 3, France

 **Present address: CSNSM F-91405 Orsay, France

 ***Present address: APC UMR7164 Universit\'e Paris VII F-75012, France

 ****Present address: Bhabha Atomic Research Center, Mumbai, India

\end{document}